\documentclass[trackchanges]{aastex62}
\received{}
\revised{}
\accepted{}
\submitjournal{ApJ}

\shorttitle{A Survey of Changes in Helicity Flux During Flares}
\shortauthors{Bi et al.}

\begin{document}

\title{A Survey of Changes in Magnetic Helicity Flux on the Photosphere During Relatively Low Class Flares }

\correspondingauthor{Yi Bi}
\email{biyi@ynao.ac.cn}

\author{Yi Bi}
\affil{ Yunnan Observatories, Chinese Academy of Sciences, 396 Yangfangwang, Guandu District, Kunming, 650216, P. R. China}
\affiliation{University of Chinese Academy of Sciences, 19A Yuquan Road, Shijingshan District, Beijing 100049, China}
\affiliation{Center for Astronomical Mega-Science, Chinese Academy of Sciences, 20A Datun Road, Chaoyang District, Beijing, 100012, P. R. China}
\affiliation{State Key Laboratory of Space Weather, Chinese Academy of Sciences, Beijing 100190}

\author{Ying D Liu}
\affiliation{State Key Laboratory of Space Weather, Chinese Academy of Sciences, Beijing 100190}

\author{Yanxiao Liu}
\affil{ Yunnan Observatories, Chinese Academy of Sciences, 396 Yangfangwang, Guandu District, Kunming, 650216, P. R. China}
\affiliation{University of Chinese Academy of Sciences, 19A Yuquan Road, Shijingshan District, Beijing 100049, China}
\affiliation{Center for Astronomical Mega-Science, Chinese Academy of Sciences, 20A Datun Road, Chaoyang District, Beijing, 100012, P. R. China}

\author{Jiayan Yang}
\affil{ Yunnan Observatories, Chinese Academy of Sciences, 396 Yangfangwang, Guandu District, Kunming, 650216, P. R. China}
\affiliation{Center for Astronomical Mega-Science, Chinese Academy of Sciences, 20A Datun Road, Chaoyang District, Beijing, 100012, P. R. China}

\author{Zhe Xu}
\affil{ Yunnan Observatories, Chinese Academy of Sciences, 396 Yangfangwang, Guandu District, Kunming, 650216, P. R. China}
\affiliation{University of Chinese Academy of Sciences, 19A Yuquan Road, Shijingshan District, Beijing 100049, China}
\affiliation{Center for Astronomical Mega-Science, Chinese Academy of Sciences, 20A Datun Road, Chaoyang District, Beijing, 100012, P. R. China}

\author{Kaifan Ji}
\affil{ Yunnan Observatories, Chinese Academy of Sciences, 396 Yangfangwang, Guandu District, Kunming, 650216, P. R. China}
\affiliation{Center for Astronomical Mega-Science, Chinese Academy of Sciences, 20A Datun Road, Chaoyang District, Beijing, 100012, P. R. China}



\begin{abstract}

Using the 135-second cadence of the photospheric vector data provided by the  Helioseismic and Magnetic Imager telescope on board the {\it Solar Dynamic Observatory}, we examined the time-evolution of  magnetic helicity fluxes across the photosphere during 16 flares with the energy class lower than M5.0.
During the flare in 4 out of 16 events, we found  impulsive changes in the helicity fluxes. 
This indicates that even the flare with less energy could be associated with anomalistic transportation of the magnetic helicity across the photosphere.
Accompanying  the impulsive helicity fluxes, the poynting fluxes across the photosphere evolved from positive to negative. As such, the transportations of magnetic  energy  across the photosphere were toward solar interior during these flares. 
In each of the  4 events,  the impulsive change in the helicity flux was always mainly contributed by abrupt change in horizontal velocity  field on a sunspot located near the flaring polarity inversion line.
The velocity field on each sunspot shows either an obvious vortex patten or an shearing patten relative to the another magnetic polarity,  which tended to relax the magnetic twist or  shear in the corona.
 During these flares, abrupt change in the Lorentz force acting on these sunspots were found. 
 The rotational motions and shearing motions of these sunspots always had the same directions with the  resultant Lorentz forces.
 These results support the view that the impulsive helicity transportation during the flare could be driven by the change in the Lorentz force applied on the photosphere.
 


\end{abstract}


\section{Introduction} \label{sec:intro}


 

The rising number of observations supports that abrupt and irreversible changes in the longitudinal  magnetic field do occur during some solar flare \citep{Patterson81,
Kosovichev99,
Kosovichev01,
Cameron99,
Sudol05,
Wang06,
Wang11,
Johnstone12,
Gosain12,
Cliver12,
Burtseva13,
Castellanos18}.
Field changes were often significantly stronger for X-class than for M-class flares \citep{Petrie10}. The impulsive field changes were also found in some C-class flare \citep{Wang13,Jing14,Castellanos18}. 
 Based on the vector data, it has been widely reported the flare-induced enhancement in the horizontal magnetic field around the flaring polarity inversion line (PIL)
 \citep{Wang05,
Wang07,
Wang10,
Li11,
Su11,
Liu12,
Petrie12,
Petrie13,
Liu13,
Song16,
Sun17,
Xu17,
Gomoy17,
Wang17}, 
which is accompanying with the impulsive enhancement in the shear angle of the magnetic field \citep{Wang94,Zhang94}. 
The increase in magnetic shear seems to contradict with the energy release for energizing flares. 
  The results of NLFFF model show that  the magnetic shear and magnetic energy in an area around the flaring PIL increased from the photosphere boundary to an altitude of $\sim$ 10 Mm, but decreased  above this space \citep{Jing08,Liu12b,Sun12}.  These results suggested that enhancements in the magnetic shear only occurred in a local area.

The enhancements in both the horizontal magnetic field and magnetic shear are often regarded as a result of the contraction of the highly sheared field produced by the coronal magnetic reconfiguration in the period of the flare \citep{Hudson00,Ji07,Hudson08}. 
  Different from the coronal loop contractions at the periphery of active regions \citep{Liu09,Sun12,Shen12,Simoes13,Kushwaha15,Wang18}, the contraction of the magnetic field around the  flaring PIL is discussed in some specific magnetic configurations.
  \citet{Wang10} proposed that the increase in the horizontal field around PIL may related to the  low-lying shorter loop across the PIL, which produced by the  the near-surface reconnection between the two sigmoid elbows as suggested by the tether-cutting model for solar eruption \citep{Moore01,Sterling03,Aulanier10,Shibata11,Chen11,Schmieder15,Chen14}.
   As suggested by \citet{Melrose12}, magnetic reconnection between current-carrying magnetic loops  can lead to   release of the stored energy, and then a net shortening of the current path, which is consistent with an increase in the horizontal component of the photospheric field during a flare. 
   \citet{Bi16} noted that the changes in the photospheric field covered by the flare ribbon may relate to the difference between the magnetic field before the flare and the newly formed field outlined by the post-flare loop.   
    
    
    Magnetic helicity and magnetic energy can be transported between the solar interior and the corona by  the motions of magnetic flux on the photosphere \citep{Berger84,Wang96,Demoulin03,Demoulin07}. 
    The rate of helicity and energy transportation  across the photosphere is defined as  helicity flux ($\dot{H}$) and energy flux $\dot{E}$, which respectively     
     can be derived as \citep{Kusano02,Liu12}
\begin{equation}
 \dot{H}=\underbrace{2\int_{S}(\mathbf{A}_{p}\cdot \mathbf{B}_{t})V_{\bot n}dS}_{\dot{H}_{e}}
 -\underbrace{2\int_{S}(\mathbf{A}_{p}\cdot \mathbf{V}_{\bot t})B_{n}dS}_{\dot{H}_{s}}
\end{equation}
and
\begin{equation}
 \dot{E}=\underbrace{\frac{1}{4\pi}\int_{S}B_{t}^{2}V_{\bot n}dS}_{\dot{E}_{e}}
 -\underbrace{\frac{1}{4\pi}\int_{S}(\mathbf{B}_{t}\cdot \mathbf{V}_{\bot t})B_{n}dS}_{\dot{E}_{s}}
\end{equation}
 where $\textbf{A}_{p}$  is the vector potential of the potential field, $\mathbf{B}$ and $\mathbf{V}_{\bot}$ denotes the magnetic field and plasma velocity perpendicular to the magnetic field, and the subscript ``t'' and ``n'' refers to the horizontal and vertical component with respect the photosphere, the subscript ``e'' and ``s'' refers to the emerging and shear term, which is contributed from shuffling horizontal motion ($\mathbf{V}_{\bot n}$) of photospheric flux  and flux emergency ($\mathbf{V}_{\bot t}$).     
 The shear term is further decomposed into the contributions from the shearing motion between the different flux patches and the spin motion of the isolated flux patch, such as the rotation of a sunspot \citep{Longcope07}. 
    Since the magnetic helicity cannot dissipate in the corona, the  helicity injected into the corona will be accumulated in the corona. 
    A mountain of  helicity accumulation is found before the flare \citep{Yamamoto05,
LaBonte07,
Park08,
Park10a,
Ravindra11,
Tziotziou13,
Guo13}.

 \citet{Moon02}  reported that the abrupt change in helicity flux  during the flare and that the impulsive helicity flux tended to have the sign opposite to that of the active region.
     \citet{Smyrli10} investigated helicity flux in the 10 active regions and found that the abrupt change in helicity flux present during the 6 flare.
    On the other hand, some works found the helicity flux changed its sign  around the start of the eruption\citep{Zhang08,Park10b,Vemareddy12,Wang14b,Gao18}, and thus the authors suggested that 
        the interaction between the magnetic flux tubes with opposite sign of helicity is fundamental for the eruption to occur \citep{Linton01,Kusano04,Liu04,Liu07,Chandra10,Romano2011a,Romano2011b}.
        Therefore, the high-cadence magnetogram is essential to  study whether the sudden change in the helicity flux is produced by the flare or vice versa.
        
          Case studies  revealed that  the impulsive change in helicity flux was mainly contributed by the abrupt reversal of rotation in a sunspot \citep{Bi16} or the   sudden rotational motion of  sunspots on both
sides of the PIL \citep{Bi17}. 
 The abrupt change in the  rotational speed of sunspots during flare was first reported by \citet{Wang14}. \citet{Liu16} noted that the sudden rotation of a sunspot occurred when the flare ribbon propagates towards the sunspot.
   The rotational motion of a sunspot can play an important role in twisting, energizing, and destabilizing the coronal magnetic field system (\citet{Fan09,Torok13}).
   It has been widely reported  that the continual rotational motions of sunspots were  followed by flare activities \citep{Brown03,
Romano05,
Regnier06,
Zang07,
Yan07,
Li09,
Min09,
Kazachenko09,
Suryanarayana10,
Jiang12,
Zhu12,
Kumar13,
Ruan14,
Bi15,
Suryanarayana15,
Li15,
Wang16,
Vemareddy16,
Yan18}.
    
    The abrupt  overall  motion of the sunspot  associated with flare \citep{Anwar93,Xu17} could be an alternative mechanism for the impulsive  change in the helicity flux.  \citet{Wang06} have reported that the sudden motions of the two magnetic polarities during flares had a tendency   to reduce the magnetic shear in the corona. The sudden shear-relaxing motion could then generate an impulse  helicity flux    having the sign opposite with that of the active region as reported by \citet{Moon02}.
       
    
    Using the vector data provided by the Helioseismic and Magnetic Imager \citep[HMI;][]{Schou} telescope on board the {\it Solar Dynamic Observatory (SDO)}, in this article, we surveyed the evolutions in  the magnetic helicity across the 16 flaring active region. During 4 out of 16 flares, significant changes in the helicity flux were found, which mainly were contributed by the abrupt motions of the  small-sized sunspots nearby the flaring PIL.  

\section{Data}

The full-disk vector magnetogram data set with 135-second cadence and a pixel size of 0.5 is used to investigate the time-evolution of the photospheric magnetic field.
Now, the 135-second cadence data provided by the HMI team \citep{Hoeksema14,Liu16b} covers some  active regions since mid 2010.
The vector field is computed using the Very Fast Inversion of the Stokes Vector code \citep{Borrero} is used to invert the vector magnetogram from a full set of Stokes
parameters (I, Q, U, V).  The remaining $180^{\circ}$ azimuth ambiguity in the vector data is resolved with the Minimum Energy algorithms \citep{Metcalf,Leka}.
The SDO/HMI also provides the full-disk continuum intensity images with a pixel size of 0 5.
 The full-disk extreme ultraviolet and  ultraviolet images with a pixel size of 0.6 from the  Atmospheric Imaging Assembly \citep[AIA;][]{Lemen} Telescopes onboard {\it SDO}  show the chromospheric and coronal structures.
 Plasma velocity $\mathbf{V}$ is estimated from the difference between the two sets of the vector data by  a differential affine velocity estimator for vector magnetograms \citep[DAVE4VM;][]{Schuck08}
   .

\begin{deluxetable*}{rlCc}[h]
\tablecaption{A list of flares surveyed.}
\tablecolumns{5}
\tablenum{1}
\tablewidth{0pt}
\tablehead{
\colhead{Number} &
\colhead{Flare} &
\colhead{NOAA} & \colhead{Name} 
}
\startdata
1 &  SOL2011-10-02T00:37(M3.9) &    11305 &    \\
2 & SOL2011-11-15T12:30(M1.9) &    11346 &    I\\
3 & SOL2011-12-27T04:11(C8.9) &   11386 &    \\
4 & SOL2012-01-19T13:44(M3.2)  &  11402 &    \\
5 & SOL2012-03-14T15:08(M2.8) &  11432 &    \\
6 & SOL2013-05-16T21:36(M1.3) &  11748 &    \\
7 & SOL2013-12-28T17:53(C9.3)  &  11936  &  II \\
8 & SOL2014-01-31T15:32(M1.1)  &  11968 &    \\
9 & SOL2014-02-01T07:14(M3.0) &   11967 &    \\
10 & SOL2014-02-12T03:52(M3.7) &  11974 &   \\
11 & SOL2014-06-12T19:56(M1.1) &  12089 &     \\
12 & SOL2014-08-01T17:55(M1.5)  &  12127 &    \\
13 & SOL2014-08-25T14:46(M2.0) &  12146 &     \\
14 & SOL2014-08-25T20:06(M3.9) &  12146 &      III\\
15 & SOL2015-11-04T13:31(M3.7) &  12443 &    \\
16 & SOL2015-11-09T23:49(M3.9) &  12449 &    IV  \\
\enddata
\end{deluxetable*}

\section{Results}
We surveyed the events that satisfy the following criteria: 
(1) HMI vector data with 135-second cadence is available spanning pre- and post-flare states for at least 2 hours;
(2) the source active region is within $60^{\circ}$ from the disk center;
(3) no M5.0-class or above flare  occurred in these regions during their disk passage.
(To avoid the ARs that are usually flare-productive  and produced many flares with various energy classes. It is interesting  to study the differences  between the changes in the helicity flux produced by the homologous flares, but it is beyond the scope of this paper.)
With these criteria, a total of 16 events are selected and are listed in Table 1.

After investigating the evolution of the magnetic helicity evolution in these events, we found during
 4 out of the 16 flares that  significant helicity flux variation occurred near the strong-field strong-sheared polarity inversion line ($PIL_{SS}$) as defined by \citet{Falconer01}.
 In this study, we label the 4 flare events as I, II, III, and  IV (see table 1).
 
The detail of these 4 events are shown in Figure 1.
     The  second row and third row of Figure 1   shows  2-hour averaged helicity flux before the onset of each flare and 9-minute averaged helicity flux during each flare, respectively.
From these panels we can see that, in the period of each flare, the abrupt change in the helicity flux occurred on a region (as the     magenta curves outlined) around the $PIL_{SS}$.
First row of Figure 1 shows that each outlined region corresponds to a distinct  magnetic flux concentration \citep{Longcope07}, which respectively covers a sunspot observed on the HMI intensity images (fourth row of Figure 1).
From Event I to IV, the area (in the units of millionth of hemisphere, $\mu$Hem) of the sunspot is 17, 15, 57, and 45 $\mu$Hem, respectively. 
The AIA/1600 \AA\ images show that the outlined regions were covered by the flare ribbons (fifth row in Figure 1), while the AIA 131 \AA\ images show that the regions were located on the endpoints of the post-flare loops (sixth row in Figure 1).

 The dark curves on  the second row of Figure 2 show   the time evolution of  helicity flux contributed from these sunspots. The helicity flux always showed an impulsive change during all of the 4 flares.
Before and after each flare, the helicity flux was always negative. However, the helicity flux become positive during flare events I, III and IV.   During Flare II,  the helicity flux across this sunspot came  close to be 0, which is consistent with both   positive and negative helicity fluxes appeared on this sunspot, as shown in the map of helicity flux (Third row of Figure 1).

The obvious changes in the energy flux were also found across these sunspots. 
As indicated by the dark curve on third row of Figure 2, the energy flux  was positive at the most of time but always became negative  in the course of each flare.

As shown on the second and third rows of Figure 2, both the shear terms (blue curves) of helicity flux and energy flux were always dominated on the emergency term (orange curves).
The  emergency term also showed impulsive changes during Flare  II and III, but no significant change was found in the period of Flare I and IV.

   Since the sunspot always covered by flare ribbons during the 4 flares, the AIA 1600 \AA\ light curve on each sunspot around the flare time appears to be a single peak curve (fourth row of Figure 2).
   The fifth row of Figure 2 exhibits the result of a cross-correlation analysis of the temporal profiles of the 1600 \AA light curve and the helicity flux contributed by the  sunspot. These panels indicate that both are correlated, with the
    largest correlation coefficient being 0.5, 0.35, 0.55, and 0.42 for each event. The correlation coefficient between the 54 pairs of random data points is about 0.34 with a probability of less than 0.01, suggesting that the found correlations are highly significant. The best correlations occur when the time-delay is 0, indicating that no time-delay greater than 135 seconds is found between the two time-evolution for each event.
    Thus, the rotational rate of the sunspot changed in a synchronous manner with 1600 \AA\ light curve from the region of sunspot.


  The first row of Figure 3 presents the DAVE4VM horizontal velocity field on each active region during each flare.  It can be seen that the velocity fields across the PILs did not show a similar pattern in the various events, but velocity fields on the sunspot  near the PIL showed a clockwise patten during Flare II, III, and IV.
Based  on DAVE4VM horizontal velocity $V_{h,i}$ at each pixel on the magnetogram, we can estimate  rotational velocity            $\omega =\frac{1}{n}\sum_{i}(\frac{\mathbf{x}_{i}- \mathbf{x}_{0}}{(\mathbf{x}_{i}- \mathbf{x}_{0})^2} \times \mathbf{V}_{h,i})$, where    $\mathbf{x}_{0}$ is the sunspot centre, n is the total number of pixels in each sunspot. 
Second row of Figure 3 shows that $\omega$ changed impulsively from positive to negative  during Flare II, III, and IV, indicating that the directions of these sunspots' rotations were reversed during these 3 flares. This suggested that the abrupt reversals of rotations in these sunspots played a role in the impulsive change in the helicity flux across these sunspots.

      The overall  translational velocity $V_{h}$  on a sunspot is  decomposed into a  Cartesian coordinate ($\hat{\mathbf{x}}, \hat{\mathbf{y}}$). Here $\hat{\mathbf{x}}$ and $\hat{\mathbf{y}}$ is parallel and perpendicular to the segment of the PIL that is located nearby the sunspot, respectively.
      Consider that these active regions have negative helicity, we define the direction of $\hat{\mathbf{x}}$ such that the movement of sunspot toward $+\hat{\mathbf{x}}$-direction would further shear the dipole over the PIL and then result in the further injection of negative helicity into the corona. Inversely, the motion toward $-\hat{\mathbf{x}}$-direction means a release of the magnetic shear. 
  The motion toward $+\hat{\mathbf{y}}$-direction makes the sunspot converge toward the PIL. 
   The fourth row of Figure 3 shows the time-evolutions of $V_{x}=\sum_{i}\mathbf{V}_{x,i}$ and $V_{y}=\sum_{i}\mathbf{V}_{y,i}$, where $i$ indicates the pixels in each sunspot. Around Flare II, the temporal profile of $V_{x}$ seem to be oscillated although the $V_{x}$ became negative during this flare.
    During Flare IV, the $V_{x}$  decreased to approximately    0. 
 During Flare I and III, it can be clearly seen that the sunspot showed a sudden  motion toward $-\hat{\mathbf{x}}$-direction, indicating a sudden  shear-relax motion, which may play a role in changing the sign in the helicity flux across the sunspot.  
      On the other hand, 
     the time-evolutions of  $V_{y}$ (Sixth row of Figure 3) did not show an obvious change in $V_{y}$ in the period of each flare.
      
      The first and second  row of Figure 4 show the spatial maps of $\delta |\mathbf{B}_{h}|$ and  $\delta |B_{r}|$, respectively. 
Each difference map was constructed by subtracting the  image at the onset of flare from the image after the peak time of flare. 
     On the most area of each sunspot, $|B_{r}|$ appeared to increase.  $|\mathbf{B}_{h}|$ increases near the $PIL_{SS}$, around which  $|\mathbf{B}_{h}|$   decreased. Both increase and decrease in $|\mathbf{B}_{h}|$ appeared on each sunspot.
     The time-evolution of averaged $B_{r}$ on each sunspot (Third row of Figure 4) shows that no significant change in $B_{r}$ occurred  during each flare. 
   The averaged   $\mathbf{B}_{h}$ on each sunspot is also decomposed into $B_{x}$ and $B_{y}$ in a cartesian coordinate such that $B_{x}$ and $B_{y}$ is  parallel and perpendicular  to the main PIL, respectively. 
    As the fifth row of Figure 4 show, the evolution of  $B_{y}$  did not show same patten during various flares;  $B_{y}$ shows a permanent decrease during Flare I and II,  a transient increase during Flare III,  and a permanent increase during Flare IV.
    The temporal profile of $B_{x}$ around Flare II (Fourth row  of Figure 4) seems oscillating and then it is difficult to determine whether the change in $B_{x}$ is related to the flare II.
     During Flare I, III, and IV, however, $B_{x}$ always shows a rapid and an irreversible increase (Fourth row  of Figure 4).
 A step function introduced by \citet{Sudol05} fits well the time-evolutions in $B_{x}$ around the times of these flare (see the red curve in Figure 4). These fitting parameters indicate that the amplitude of the impulsive change in $B_{x}$ amounts to 62G, 48G, and 39G in the period of Flare I, III, and IV, respectively.

         Following \citet{Fisher12}, the volume integral of the Lorentz force can be written as 
             \begin{eqnarray}
             \mathbf{F}_{L}=\frac{1}{8\pi}\oint_{S}d^{2}x[2\mathbf{B}(\mathbf{B}\cdot \hat{\mathbf{n}})-\hat{\mathbf{n}}B^{2}],
\end{eqnarray}
where S represents the area of the entire
bounding surface.
      The authors considered that the volume of an active region is observed to be static and then the total force acted to an active region must be close to 0 when the eruption is absent;
      Moreover, the authors  assumed that the magnetic field integrated over the upper surface and side walls of the volume above the active region is negligible. Based on these assumptions, the Lorentz force acted on the photosphere must be balanced  by other forces such as gas-pressure gradients and gravity.
      The abrupt change in the magnetic field on the photosphere would produce an abrupt change in the Lorentz force, which could then produce an imbalance in the photosphere until a new equilibrium is reached. According to Equation (3),    the  change in the horizontal component of the the Lorentz force  applied on the photosphere is 
       \begin{eqnarray}
\delta\mathbf{F}_{h}=\frac{1}{4\pi}\int_{S}d^{2}x\delta[B_{r}\mathbf{B}_{h}]
\end{eqnarray}
Here, $F_{h}$ is decomposed into $F_{x}$ and $F_{y}$ in the coordinate as above.  The time-evolution of $\delta F_{x}$ (Fifth row of Figure 3) shows that all of the sunspots were abruptly exerted by a force toward the direction that relax the shear of coronal magnetic, which is consistent with the overall velocity of the sunspot during  Flare I and III.
      Moreover,   $\delta F_{y}$ (Seventh row of figure 3) applied on the sunspot also shows an abrupt change toward +y-direction, which may drive the sunspot converge toward PIL.  However,   no significant converging velocity was found during each flare. 
      
 If S indicates the area of a sunspot, then the $\delta\mathbf{F}_{h}$ corresponds to the change in horizontal force acted on the sunspot.
The change in the torque acted on the sunspot then can be written as
       \begin{eqnarray}
\delta\mathbf{T}=\frac{1}{4\pi}\int_{S}d^{2}x (\mathbf{x}-\mathbf{x}_{0}) \times \delta[B_{r}\textbf{B}_{h}]
\end{eqnarray}     
            A downward torque $\mathbf{\delta T}$ can be found during the flare in each event (Third row of Figure 3). This is consistent with abrupt clockwise rotation of the sunspot observed during Flare II, III, and IV,

 Using the HMI vector data as the boundary condition and the potential field as the initial condition, one can model the magnetic field in the corona based on the force free assumption. The force-free field means that $\nabla \times \mathbf{B}=\alpha\mathbf{B}$ and thus $\mathbf{B}\cdot\nabla\alpha=0$, which indicate that the value of $\alpha$ along a force-free field line is constant. In this study, a nonlinear force-free (NLFFF) extrapolation code based on the optimization method \citep{Wheatland00,Wiegelmann04} is used to model the coronal magnetic field $\mathbf{B}$.
 
 Figure  5 shows scatter plots of the values of the $\alpha$ versus the strengths of the modeled magnetic field $\mathbf{B}$. Here, the field strength  is normalized to unity in the peak value of the  field strength along a magnetic field line.
 From these plots one can  see  that the values of $\alpha$ measured in the weak field (lower than 0.5) have much larger divergence than that in the strong field (greater than 0.5).
 However, the value of $\alpha$ should be constant along a force-free field line. It thus seems that the value of $\alpha$ derived numerically in the weak field should be unphysical. 
 For this reason, we estimate the value of $\alpha$ of a modeled field line by using a  weighted average of $\alpha$, the weight  depending on $B^4$, which ensures that the weight factor for the weak field is less than 10\% of that for the strong field.
For a field line, the magnetic-weighted average of value of $\alpha$ is defined as 
  \begin{equation}
  \bar{\alpha}=\frac{\sum_{i} B_{i}^{4} \alpha_{i}}{\sum_{i} B_{i}^{4}}
  \end{equation}
  where $\alpha_{i}=(\nabla \times \mathbf{B}_{i}) \cdot \mathbf{B}_{i}/\mathbf{B}^{2}_{i}$ , and the subscript i runs over all the sampled points on a field line.

  The first row of Figure 6 shows the maps of  $\bar{\alpha}$. In each event, the negative value of $\bar{\alpha}$ is dominated around the $PIL_{SS}$. Also, the values of $\bar{\alpha}$ in each region covering the sunspot are almost negative. The second row of Figure 6 shows the difference between the maps of  $|\bar{\alpha}|$  before and after the flare, from which we can see that the value of  $|\bar{\alpha}|$  in the most area around the $PIL_{SS}$ decreased, but  enhanced in the sunspots during the flare.  The third row of Figure 6 shows the time-evolution of  averaged value of $\bar{\alpha}$ of the field lines that are traced from the sunspots outlined. It can be clearly see that the enhancement in the value of $|\bar{\alpha}|$ during each flare is impulsive and irreversible.
  Meanwhile, the averaged lengths of the field lines starting from the sunspots show a impulsive decrease, indicating that the field lines were shortened during the flare (Fourth row of Figure 6).
Again, a stepwise function fits well the time-evolution of the value of $\bar{\alpha}$ and the length of the field lines.
The best-fitting parameters show that the amplitude of impulsive increase in $|\alpha|$ is 0.007/Mm, 0.015/Mm, 0.003/Mm, 0.02/Mm during Flare I, II, III, and IV, respectively. The length of the field lines decreased impulsively by  3 Mm, 8 Mm, 12 Mm, and 4 Mm, for each event, respectively

      
       

  \section{Discussions}
  After surveying 16 flares with class lower than M5.0, we found abrupt change in the helicity flux in the 4 events. This indicates that even the flare with less energy could occur with an
  anomalistic transportation of the magnetic helicity across the photosphere. During each flare of the 4 flares,  same similarity as follow:
  \begin{enumerate}
 \item  The impulsive change in the helicity flux was mainly contributed by either the shear or rotational motion of a sunspot,
 which was located near the $PIL_{SS}$ and was covered by flare ribbons. 
 \item The helicity flux across the sunspot   had the sign opposite with that of the accumulated helicity before the flare.
\item The energy flux across the sunspot  was  negative during the flare while positive before the flare.
\item The photospheric magnetic field on the sunspot showed an abrupt and irreversible changes.  
\item The changes in the Lorentz force  always tended to rotate these sunspots to decrease the coronal magnetic twist   and move these sunspots along  the PIL to reduce the magnetic shear in the corona.  Consistently, the sudden reversals in the rotations of the sunspots were detected during Flare II, III, and IV, and the impulsive shear-reducing motions were found during Flare I and III.
\item The values of $\bar{\alpha}$ of the NLFFF field connecting the sunspot showed a rapid  and irreversible increase.
\end{enumerate}

    
    Since all of the sunspots were located on the endpoints of the flaring loop,  the sunspots were connected to the reconnected field in the course of the flares.
        This suggests that the abrupt  reversal in the rotation of the sunspot is related to the magnetic reconnection during the flare.
     This is further supported by the fact that rotation of the sunspot    show synchronous changes with 1600 \AA\  emission from the regions of the sunspots that are closest to the PIL.

 Since twisted and sheared fields store the free magnetic energy, the rotational motion of a sunspot relates to the transportation of the magnetic helicity and magnetic energy across the sunspot.
    The negative values of the energy flux across the sunspot was found during each flare reported here, indicating that the magnetic energy was impulsively transported toward  solar interior. Accordingly,  the impulsive changes in helicity flux during these flares   are  more possible to be a consequence of transpotation from solar atmosphere toward  solar interior, instead of  the opposite-signed    helicity  injected from solar interior  into the active region during these flares. 

   Both the rotational motion of a sunspot and the shear motion of sunspots on the two sides of PIL are basically  the consequence of  the propagation of a shear Alfv$\acute{e}$n wave across the photosphere. Specifically,  
 the rotational motion of a sunspot tends to remove the gradient of the value of $\alpha$ between the solar corona and the solar interior, which produces the Lorentz force that drives sunspot rotation \citep{Parker79,Longcope00,Magara03,Chae03,Fan09,Sturrock15}.
If a  sunspot rotates in such a direction that  the magnetic helicity is transported form the corona into solar interior,   the value of $\alpha$ in the corona should be higher than that in the solar interior.  Consistently,   the results of the NLFFF model  show that the value of $\alpha$ aways increased during the flare for all events.
Considering that the value of $\alpha$ in the solar interior did not change significantly, we suggest that the flare-induced enhancement in the value of coronal $\alpha$ could change a gradient of value of $\alpha$ form the solar corona to the solar interior.

The  sunspots on the two sides of PIL would be exerted by the opposite  Lorentz force    
  $\mathbf{F}_{x}=-\partial(B^{2}/2)/\partial x+(\mathbf{B} \cdot \nabla)\mathbf{B}_{x}$ \citep{Manchester00,Manchester01,Fan01,Manchester04,Archontis04,Torok14}, 
where  $\hat{\mathbf{x}}$ is  parallel to the PIL.  Similarly,  $\mathbf{F}_x$ would drive a shear motion so as to remove a gradient of $B_{x}$ along a field line.
Resulted from  the upward expansion of the a twisted flux tube,  a decrease in $B_{x}$ in the corona would result in a $F_{x}$ that tends to drive a shear motion to accumulate the magnetic shear in the coronal. In this case, this shearing process produced eruptions that are representative of coronal mass ejections and flares. During flare, a reverse process was reported here.
During Flare I and III, the unambiguous shear-relax motions of the sunspots were found to be accompanying with the impulsive increase of the horizontal field along the PIL  and the contraction of the field lines as indicated by the NLFFF model. Such an increase in the shear component of the magnetic field is found in the collapsing loop system shown in \citet{Manchester07}
 
    
  The  4 sunspots reported here have the area   smaller than 100 $\mu$Hem, which  fall into the category of small-sized sunspot as defined by \citet{Mandal16}, who defined the sunspot with the area larger than 200 uHem as the large-sized sunspot.
  It seems that the flares with the relative low energy have no enough energy to power the motion of the large-sized sunspot, but more surveys is need to clarify  how the photospheric motions produced by the flare are related to the flare energy.

\section{Acknowledgements}

    The authors are grateful to the anonymous referee for  detailed comments and useful suggestions for improving this manuscript.      This work is supported by the Natural Science
Foundation of China under grants 11633008, 11573012, 11333007, 11503081, Project Supported by the Specialized Research Fund for State Key Laboratories.
This work use  the DAVE/DAVE4VM codes developed by the Naval Research
Laboratory and  the NLFFF code written by Dr Wiegelmann. The NASA/SDO data used here are courtesy of the
 HMI and AIA science teams.            

\begin{figure}[h]
\plotone{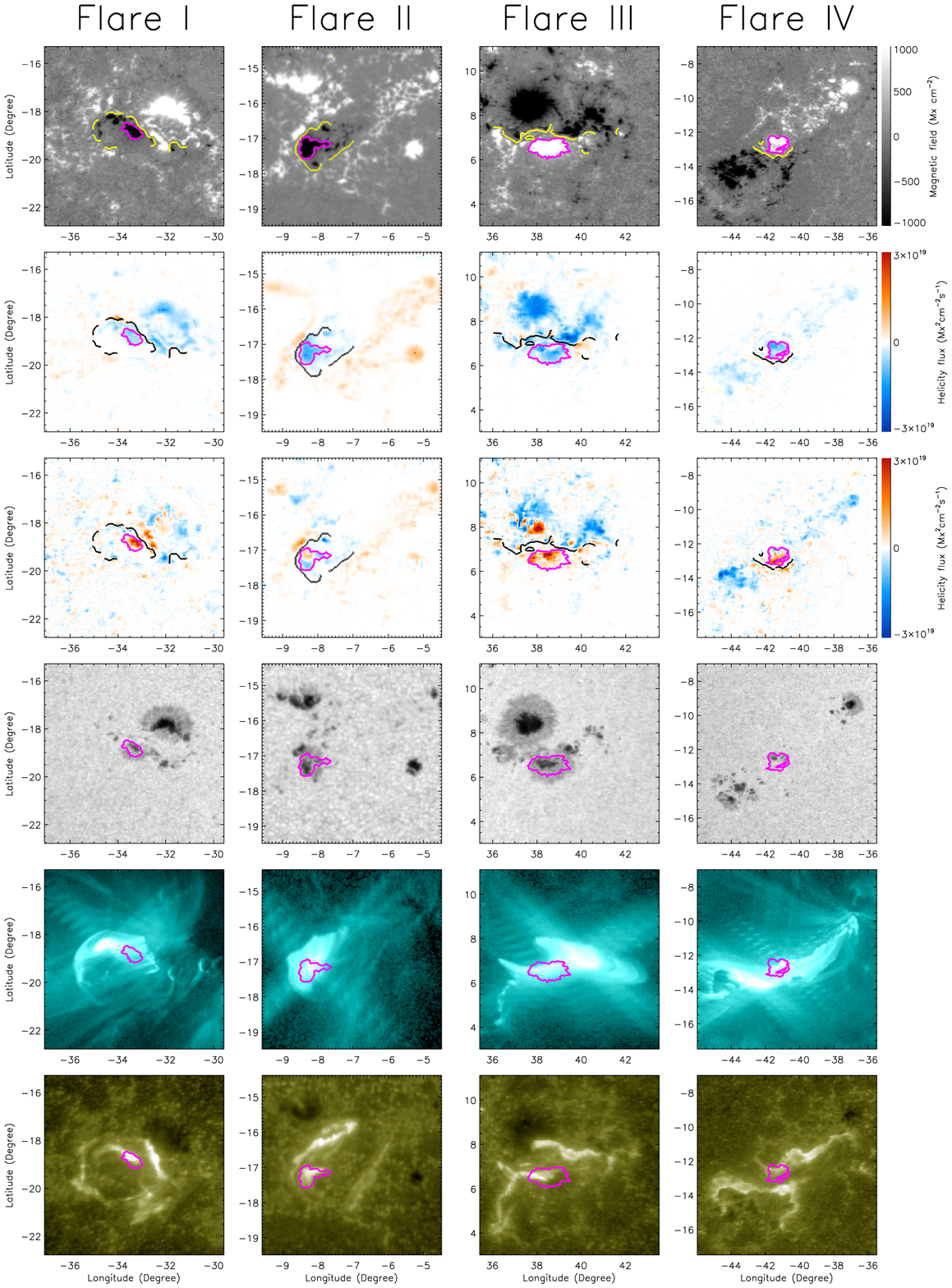}
\caption{
  First row: the vertical component of the HMI vector data. Second row:  2-hour time-average map of helicity flux before the flare.
  Third row:  9-minute time-average map of helicity flux during the flare.
   Fourth row: SDO/HMI intensity images. Fifth row: SDO/AIA 1600 \AA\ images.
   Sixth row: SDO/AIA 131 \AA\ images. In each panels, the magenta curve encloses a distinct  magnetic flux concentration, indicating a sunspot; the dark and yellow curves refer to $PIL_{SS}$. 
}
\end{figure}

 \begin{figure}[h]
\plotone{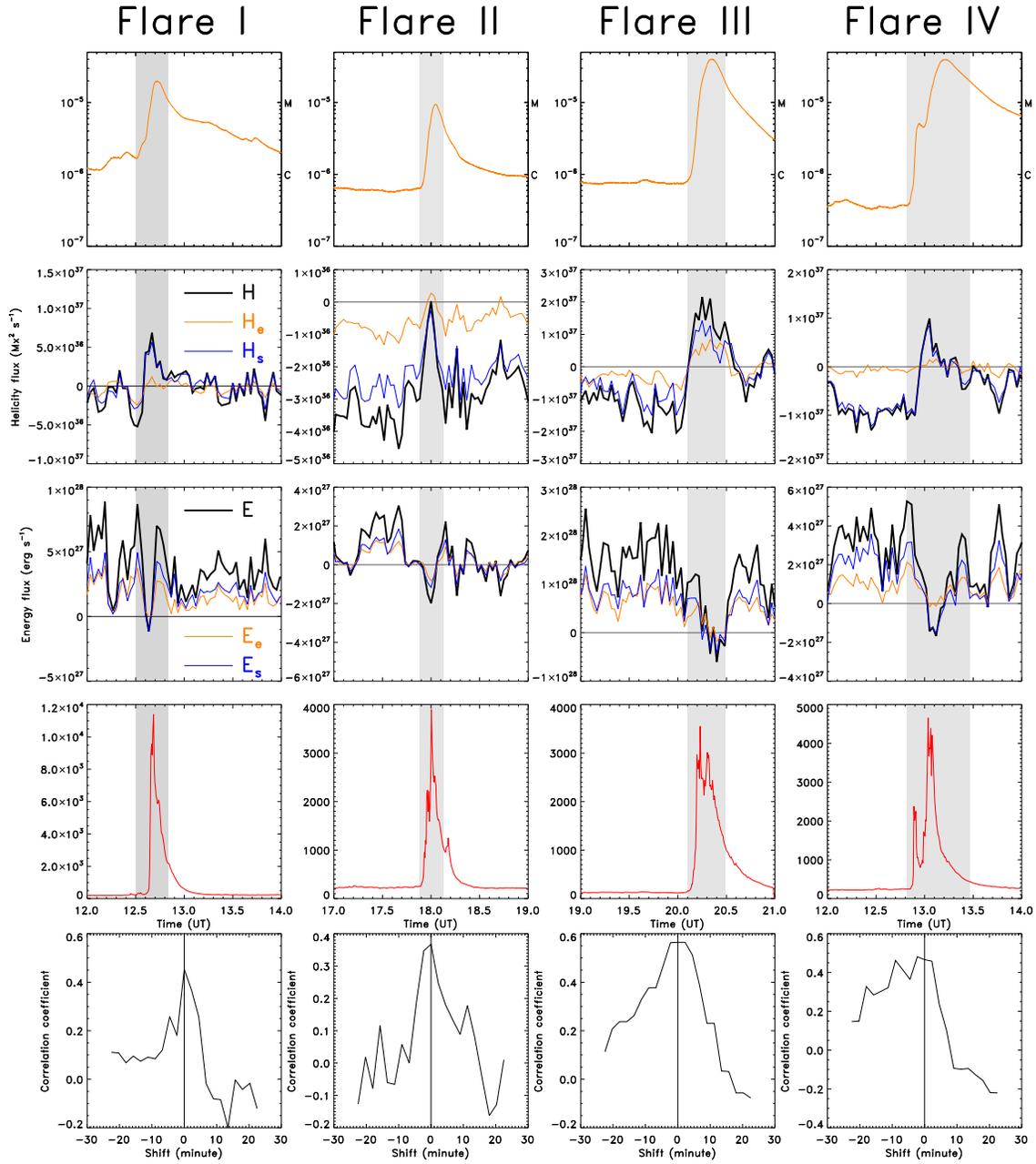}
\caption{    First row: GOES light curves.
Second row: Temporal profile of integral helicity flux contributed from each sunspot. 
Third row: Temporal profile of integral energy flux contributed from each sunspot. 
         Fourth row:   AIA 1600 \AA\ light curve on each sunspot. 
      Fifth row:  Cross-correlation coefficient between the 1600 \AA\ light curve and helicity flux  contributed from the sunspot.
       Each sunspot is defined as  the region enclosed by the red curve as the Figure 1 show.
}
\end{figure}

      \begin{figure}[h]
\plotone{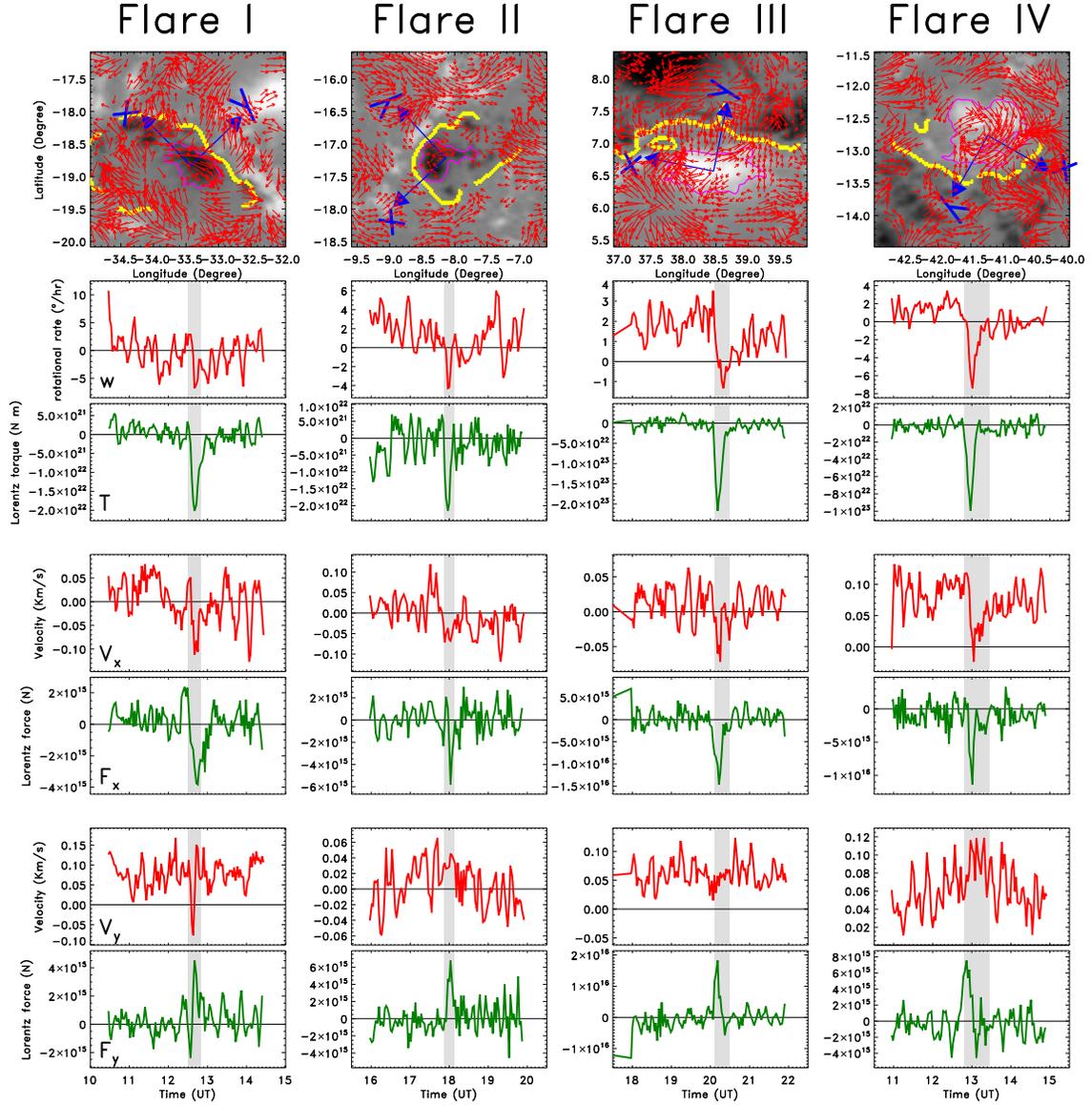}
\caption{  First row: the maps of vertical field  superimposed with tangential velocity vectors (red arrows) inferred from the DAVE4VM technique. 
 In each panels, the magenta curve encloses a distinct  magnetic flux concentration, indicating a sunspot;  yellow curves refer to $PIL_{SS}$. 
From second row to seventh row, the temporal profile shows the time-evolution of averaged rotational rate of each sunspot (second row),  the Lorentz force torque applied on each sunspot (third row), overall velocity $V_{x}$  of each sunspot (fourth row),
integral Lorentz force ($F_{x}$) acted on each sunspot (fifth row),   overall velocity $V_{y}$  of each sunspot (sixth row),
and integral Lorentz force ($F_{y}$) acted on each sunspot (seventh row). The gray band in each panel denotes GOES flare time. 
}
\end{figure}

    \begin{figure}[h]
\plotone{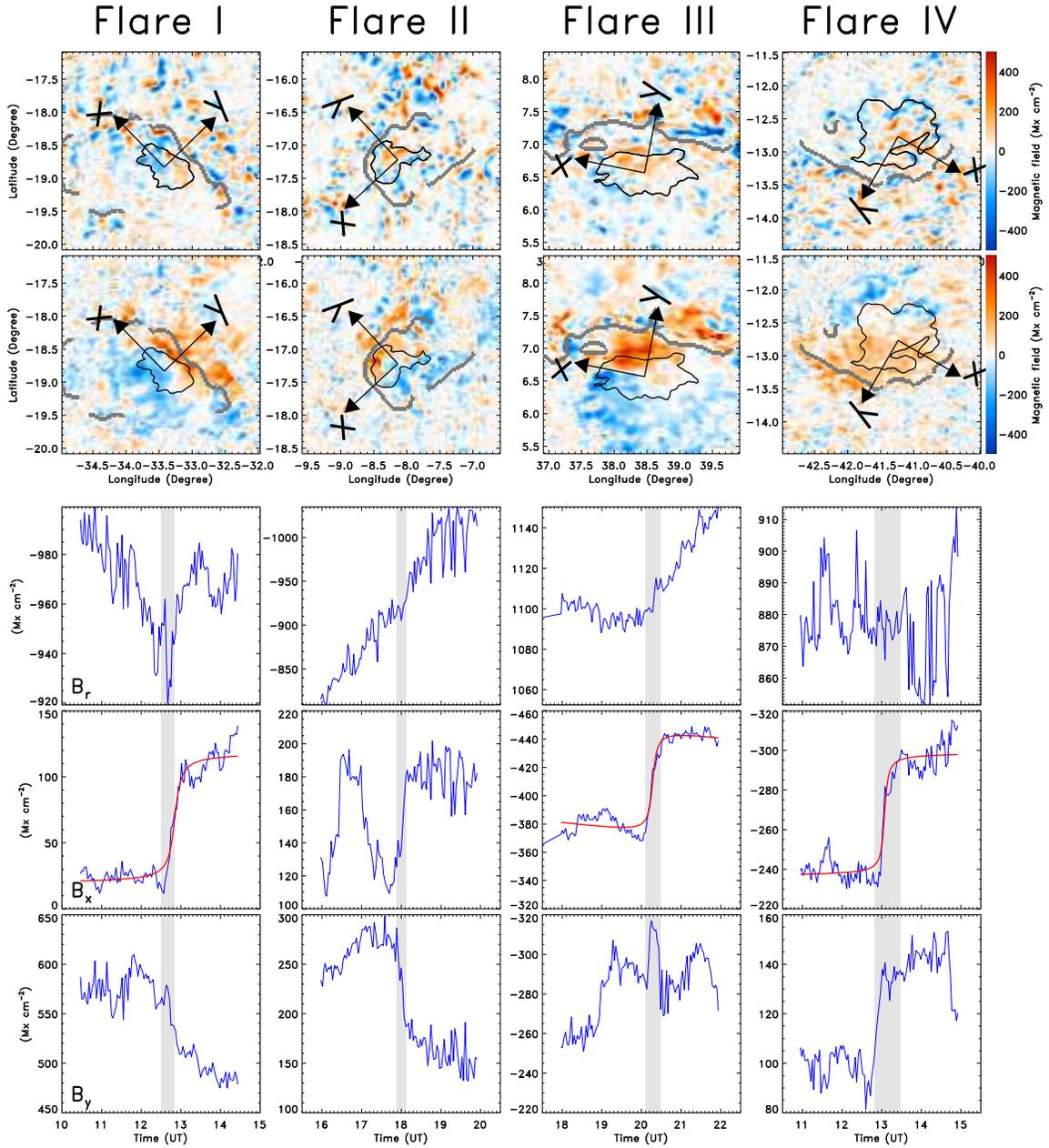}
\caption{
First row:  The difference between $|B_{r}|$ before and after flare.
Second row: The difference between $|B_{h}|$ before and after flare.
From third to fifth row, the temporal profile shows the time-evolution of averaged  $B_{r}$ (third row), $B_{x}$ (fourth row), and $B_{y}$ (fifth row) on each sunspot.
In each panel, the gray band denotes GOES flare time;
the red curves refer to the best fit of a stepwise function to the data.
}
\end{figure}

        \begin{figure}[!h]
\plotone{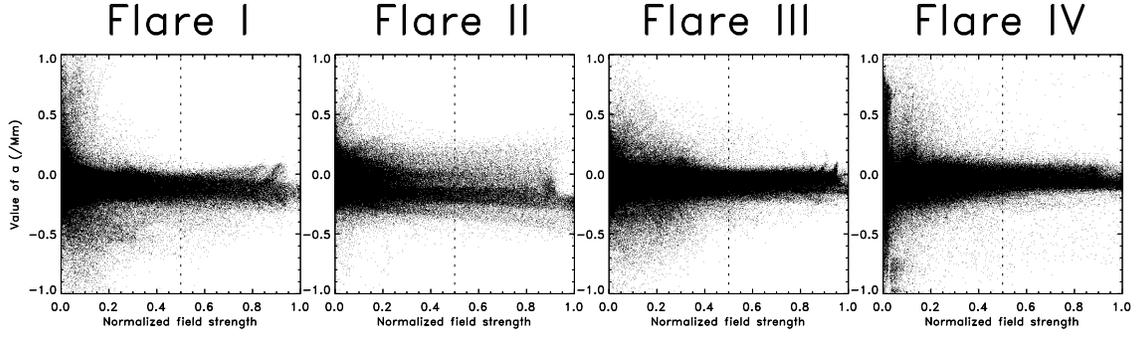}
\caption{  Scatter plots of  the value of the $\alpha$ against the normalized strength of the magnetic field, which  is normalized to unity in the peak value of the  field strength along a magnetic field line. The sampling points are located on the field lines starting from the sunspot concerned here. }
\end{figure}
            
              \begin{figure}[h!]
\plotone{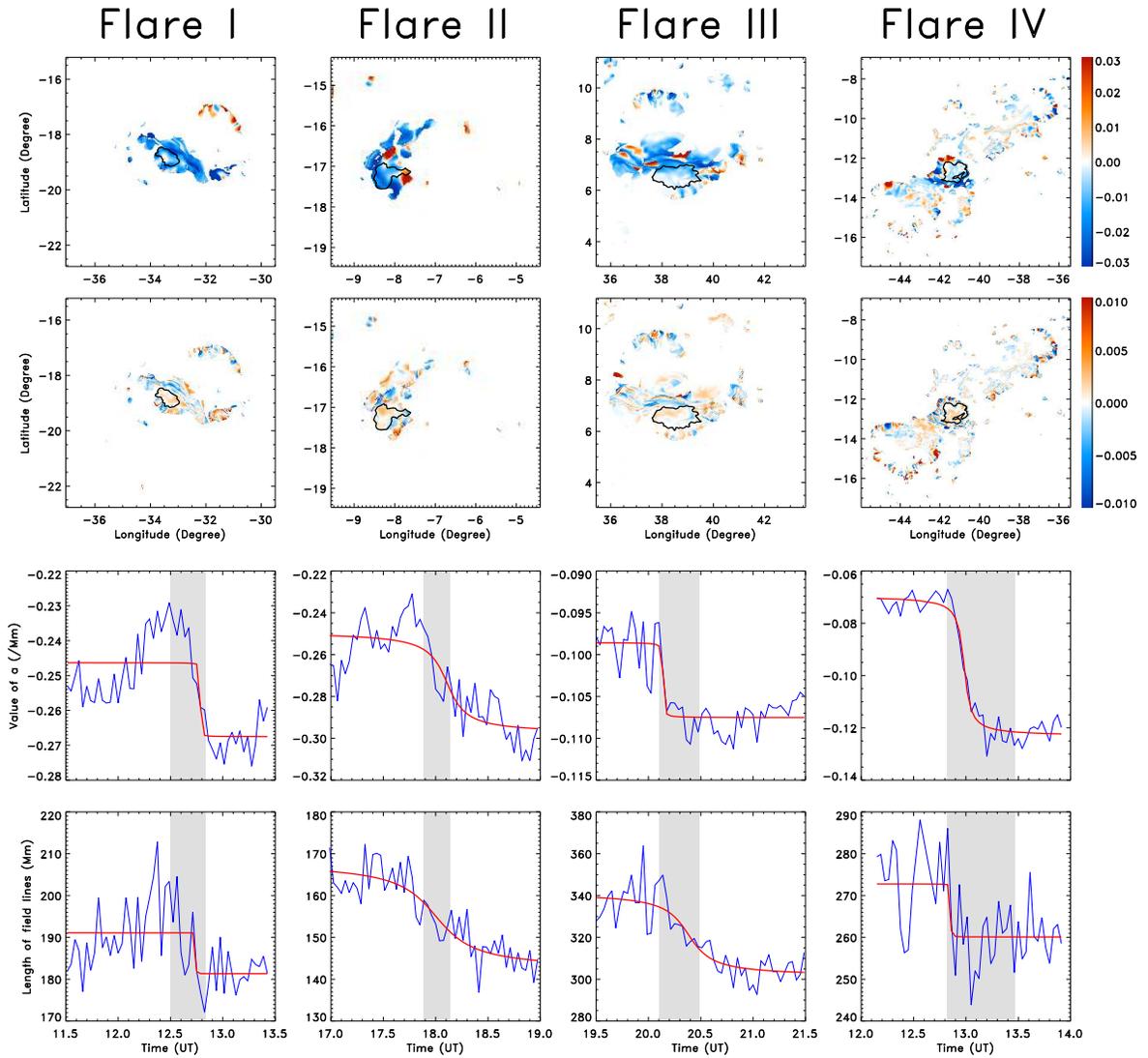}
\caption{ First row: the $\bar{\alpha}$ maps. 
  Second row: the difference between the $\bar{\alpha}$ before and after flare.
  Third row: the time-evolution of the average of $\bar{\alpha}$ of the field lines starting from the region as outlined by the red curves in Figure 1.
  Fourth row: the time-evolution of the average of length of the field lines.  In each panel on the third and fourth rows, the gray band denotes GOES flare time. 
  The red curves refer to the best fit of a stepwise function to the data.}
\end{figure}

\listofchanges


\begin{thebibliography}{}


\bibitem[Anwar et al.(1993)]{Anwar93} Anwar, B., Acton, L.~W., Hudson, H.~S., et al.\ 1993, \solphys, 147, 287 

\bibitem[Archontis et al.(2004)]{Archontis04} Archontis, V., Moreno-Insertis, F., Galsgaard, K., Hood, A., \& O'Shea, E.\ 2004, \aap, 426, 1047 


\bibitem[Aulanier et al.(2010)]{Aulanier10} Aulanier, G., T{\"o}r{\"o}k, T., D{\'e}moulin, P., \& DeLuca, E.~E.\ 2010, \apj, 708, 314 


\bibitem[Berger \& Field(1984)]{Berger84} Berger, M.~A., \& Field, G.~B.\ 1984, Journal of Fluid Mechanics, 147, 133 
\bibitem[Bi et al.(2015)]{Bi15} Bi, Y., Jiang, Y., Yang, J., et al.\ 2015, \apj, 805, 48 
\bibitem[Bi et al.(2016)]{Bi16} Bi, Y., Jiang, Y., Yang, J., et al.\ 2016, Nature Communications, 7, 13798 

\bibitem[Bi et al.(2017)]{Bi17} Bi, Y., Yang, J., Jiang, Y., et al.\ 2017, \apjl, 849, L35 
\bibitem[Brown et al.(2003)]{Brown03} Brown, D.~S., Nightingale, R.~W., Alexander, D., et al.\ 2003, \solphys, 216, 79 


\bibitem[Borrero et al.(2011)]{Borrero} Borrero, J.~M., Tomczyk, S., Kubo, M., et al.\ 2011, \solphys, 273, 267
\bibitem[Burtseva \& Petrie(2013)]{Burtseva13} Burtseva, O., \& Petrie, G.\ 2013, \solphys, 283, 429 

\bibitem[Cameron \& Sammis(1999)]{Cameron99} Cameron, R., \& Sammis, I.\ 1999, \apjl, 525, L61 
\bibitem[Castellanos Dur{\'a}n et al.(2018)]{Castellanos18} Castellanos Dur{\'a}n, J.~S., Kleint, L., \& Calvo-Mozo, B.\ 2018, \apj, 852, 25
\bibitem[Chae et al.(2003)]{Chae03} Chae, J., Moon, Y.-J., Rust, D.~M., Wang, H., \& Goode, P.~R.\ 2003, Journal of Korean Astronomical Society, 36, 33 
\bibitem[Chandra et al.(2010)]{Chandra10} Chandra, R., Pariat, E., Schmieder, B., Mandrini, C.~H., \& Uddin, W.\ 2010, \solphys, 261, 127 

\bibitem[Chen et al.(2014)]{Chen14} Chen, H., Zhang, J., Cheng, X., et al.\ 2014, \apjl, 797, L15 
\bibitem[Chen(2011)]{Chen11} Chen, P.~F.\ 2011, Living Reviews in Solar Physics, 8, 1 


\bibitem[Cliver et al.(2012)]{Cliver12} Cliver, E.~W., Petrie, G.~J.~D., \& Ling, A.~G.\ 2012, \apj, 756, 144 



\bibitem[D{\'e}moulin \& Berger(2003)]{Demoulin03} D{\'e}moulin, P., \& Berger, M.~A.\ 2003, \solphys, 215, 203 

\bibitem[D{\'e}moulin(2007)]{Demoulin07} D{\'e}moulin, P.\ 2007, Advances in Space Research, 39, 1674 


\bibitem[Falconer(2001)]{Falconer01} Falconer, D.~A.\ 2001, \jgr, 106, 25185 

\bibitem[Fan(2001)]{Fan01} Fan, Y.\ 2001, \apjl, 554, L111 


\bibitem[Fan(2009)]{Fan09} Fan, Y.\ 2009, \apj, 697, 1529 


\bibitem[Fisher et al.(2012)]{Fisher12} Fisher, G.~H., Bercik, D.~J., Welsch, B.~T., \& Hudson, H.~S.\ 2012, \solphys, 277, 59 


\bibitem[Gao(2018)]{Gao18} Gao, Y.\ 2018, Research in Astronomy and Astrophysics, 18, 028 

\bibitem[G{\"o}m{\"o}ry et al.(2017)]{Gomoy17} G{\"o}m{\"o}ry, P., Balthasar, H., Kuckein, C., et al.\ 2017, \aap, 602, A60 

\bibitem[Gosain(2012)]{Gosain12} Gosain, S.\ 2012, \apj, 749, 85 
\bibitem[Guo et al.(2013)]{Guo13} Guo, Y., Ding, M.~D., Cheng, X., Zhao, J., \& Pariat, E.\ 2013, \apj, 779, 157 






\bibitem[Hoeksema et al.(2014)]{Hoeksema14} Hoeksema, J.~T., Liu, Y., Hayashi, K., et al.\ 2014, \solphys, 289, 3483 

\bibitem[Hudson(2000)]{Hudson00} Hudson, H.~S.\ 2000, \apjl, 531, L75 
\bibitem[Hudson et al.(2008)]{Hudson08} Hudson, H.~S., Fisher, G.~H., \& Welsch, B.~T.\ 2008, Subsurface and Atmospheric Influences on Solar Activity, 383, 221 


\bibitem[Ji et al.(2007)]{Ji07} Ji, H., Huang, G., \& Wang, H.\ 2007, \apj, 660, 893 

\bibitem[Jiang et al.(2012)]{Jiang12} Jiang, Y., Zheng, R., Yang, J., et al.\ 2012, \apj, 744, 50 

\bibitem[Jing et al.(2014)]{Jing14} Jing, J., Liu, C., Lee, J., et al.\ 2014, \apjl, 784, L13 
\bibitem[Jing et al.(2008)]{Jing08} Jing, J., Wiegelmann, T., Suematsu, Y., Kubo, M., \& Wang, H.\ 2008, \apjl, 676, L81 



\bibitem[Johnstone et al.(2012)]{Johnstone12} Johnstone, B.~M., Petrie, G.~J.~D., \& Sudol, J.~J.\ 2012, \apj, 760, 29 





\bibitem[Kazachenko et al.(2009)]{Kazachenko09} Kazachenko, M.~D., Canfield, R.~C., Longcope, D.~W., et al.\ 2009, \apj, 704, 1146 
\bibitem[Kumar et al.(2013)]{Kumar13} Kumar, P., Park, S.-H., Cho, K.-S., \& Bong, S.-C.\ 2013, \solphys, 282, 503 



\bibitem[Kusano et al.(2002)]{Kusano02} Kusano, K., Maeshiro, T., Yokoyama, T., \& Sakurai, T.\ 2002, \apj, 577, 501 


\bibitem[Kusano et al.(2004)]{Kusano04} Kusano, K., Maeshiro, T., Yokoyama, T., \& Sakurai, T.\ 2004, \apj, 610, 537 

\bibitem[Kushwaha et al.(2015)]{Kushwaha15} Kushwaha, U., Joshi, B., Veronig, A.~M., \& Moon, Y.-J.\ 2015, \apj, 807, 101 



\bibitem[Kosovichev \& Zharkova(1999)]{Kosovichev99} Kosovichev, A.~G., \& Zharkova, V.~V.\ 1999, \solphys, 190, 459 

\bibitem[Kosovichev \& Zharkova(2001)]{Kosovichev01} Kosovichev, A.~G., \& Zharkova, V.~V.\ 2001, \apjl, 550, L105 
\bibitem[LaBonte et al.(2007)]{LaBonte07} LaBonte, B.~J., Georgoulis, M.~K., \& Rust, D.~M.\ 2007, \apj, 671, 955 
\bibitem[Leka et al.(2009)]{Leka} Leka, K.~D., Barnes, G., Crouch, A.~D., et al.\ 2009, \solphys, 260, 83


\bibitem[Lemen et al.(2012)]{Lemen} Lemen, J.~R., Title, A.~M., Akin, D.~J., et al.\ 2012, \solphys, 275, 17

\bibitem[Li \& Liu(2015)]{Li15} Li, A., \& Liu, Y.\ 2015, \solphys, 290, 2199 
\bibitem[Li \& Zhang(2009)]{Li09} Li, L., \& Zhang, J.\ 2009, \apjl, 706, L17 
\bibitem[Li et al.(2011)]{Li11} Li, Y., Jing, J., Fan, Y., \& Wang, H.\ 2011, \apjl, 727, L19 
\bibitem[Linton et al.(2001)]{Linton01} Linton, M.~G., Dahlburg, R.~B., \& Antiochos, S.~K.\ 2001, \apj, 553, 905 


\bibitem[Liu et al.(2012b)]{Liu12b} Liu, C., Deng, N., Liu, R., et al.\ 2012, \apjl, 745, L4 

\bibitem[Liu et al.(2013)]{Liu13} Liu, C., Deng, N., Lee, J., et al.\ 2013, \apjl, 778, L36 

\bibitem[Liu et al.(2016)]{Liu16} Liu, C., Xu, Y., Cao, W., et al.\ 2016, Nature Communications, 7, 13104 


\bibitem[Liu \& Wang(2009)]{Liu09} Liu, R., \& Wang, H.\ 2009, \apjl, 703, L23 

\bibitem[Liu \& Kurokawa(2004)]{Liu04} Liu, Y., \& Kurokawa, H.\ 2004, \pasj, 56, 497 
\bibitem[Liu et al.(2007)]{Liu07} Liu, Y., Kurokawa, H., Liu, C., et al.\ 2007, \solphys, 240, 253 
\bibitem[Liu \& Schuck(2012)]{Liu12} Liu, Y., \& Schuck, P.~W.\ 2012, \apj, 761, 105 
\bibitem[Liu et al.(2016b)]{Liu16b} Liu, Y., Baldner, C., Bogart, R.~S., et al.\ 2016, AAS/Solar Physics Division Abstracts \#47, 47, 8.10 
\bibitem[Longcope \& Welsch(2000)]{Longcope00} Longcope, D.~W., \& Welsch, B.~T.\ 2000, \apj, 545, 1089 

\bibitem[Longcope et al.(2007)]{Longcope07} Longcope, D.~W., Ravindra, B., \& Barnes, G.\ 2007, \apj, 668, 571 


\bibitem[Magara \& Longcope(2003)]{Magara03} Magara, T., \& Longcope, D.~W.\ 2003, \apj, 586, 630 
\bibitem[Manchester \& Low(2000)]{Manchester00} Manchester, W., \& Low, B.~C.\ 2000, Physics of Plasmas, 7, 1263 

\bibitem[Manchester(2001)]{Manchester01} Manchester, W., IV 2001, \apj, 547, 503 

\bibitem[Manchester et al.(2004)]{Manchester04} Manchester, W., IV, Gombosi, T., DeZeeuw, D., \& Fan, Y.\ 2004, \apj, 610, 588 
\bibitem[Manchester(2007)]{Manchester07} Manchester, W., IV 2007, \apj, 666, 532 

\bibitem[Mandal \& Banerjee(2016)]{Mandal16} Mandal, S., \& Banerjee, D.\ 2016, \apjl, 830, L33 
\bibitem[Melrose(2012)]{Melrose12} Melrose, D.~B.\ 2012, \apj, 749, 58 
\bibitem[Metcalf(1994)]{Metcalf} Metcalf, T.~R.\ 1994, \solphys, 155, 235
\bibitem[Min \& Chae(2009)]{Min09} Min, S., \& Chae, J.\ 2009, \solphys, 258, 203

\bibitem[Moon et al.(2002)]{Moon02} Moon, Y.-J., Chae, J., Wang, H., Choe, G.~S., \& Park, Y.~D.\ 2002, \apj, 580, 528 

\bibitem[Moore et al.(2001)]{Moore01} Moore, R.~L., Sterling, A.~C., Hudson, H.~S., \& Lemen, J.~R.\ 2001, \apj, 552, 833 



\bibitem[Park et al.(2008)]{Park08} Park, S.-H., Lee, J., Choe, G.~S., et al.\ 2008, \apj, 686, 1397-1403 
\bibitem[Park et al.(2010a)]{Park10a} Park, S.-h., Chae, J., \& Wang, H.\ 2010, \apj, 718, 43 
\bibitem[Park et al.(2010b)]{Park10b} Park, S.-H., Chae, J., Jing, J., Tan, C., \& Wang, H.\ 2010, \apj, 720, 1102 
\
\bibitem[Parker(1979)]{Parker79} Parker, E.~N.\ 1979, Oxford, Clarendon Press; New York, Oxford University Press, 1979, 858 p.,  


\bibitem[Patterson \& Zirin(1981)]{Patterson81} Patterson, A., \& Zirin, H.\ 1981, \apjl, 243, L99 

\bibitem[Petrie \& Sudol(2010)]{Petrie10} Petrie, G.~J.~D., \& Sudol, J.~J.\ 2010, \apj, 724, 1218 
\bibitem[Petrie(2012)]{Petrie12} Petrie, G.~J.~D.\ 2012, \apj, 759, 50 

\bibitem[Petrie(2013)]{Petrie13} Petrie, G.~J.~D.\ 2013, \solphys, 287, 415 

\bibitem[Qiu \& Gary(2003)]{Qiu03} Qiu, J., \& Gary, D.~E.\ 2003, \apj, 599, 615 

\bibitem[Ravindra et al.(2011)]{Ravindra11} Ravindra, B., Yoshimura, K., \& Dasso, S.\ 2011, \apj, 743, 33 

\bibitem[R{\'e}gnier \& Canfield(2006)]{Regnier06} R{\'e}gnier, S., \& Canfield, R.~C.\ 2006, \aap, 451, 319 
\bibitem[Romano et al.(2005)]{Romano05} Romano, P., Contarino, L., \& Zuccarello, F.\ 2005, \aap, 433, 683 
\bibitem[Romano et al.(2011)]{Romano2011a} Romano, P., Pariat, E., Sicari, M., \& Zuccarello, F.\ 2011, \aap, 525, A13 

\bibitem[Romano \& Zuccarello(2011)]{Romano2011b} Romano, P., \& Zuccarello, F.\ 2011, \aap, 535, A1 
\bibitem[Ruan et al.(2014)]{Ruan14} Ruan, G., Chen, Y., Wang, S., et al.\ 2014, \apj, 784, 165 

\bibitem[Schmieder et al.(2015)]{Schmieder15} Schmieder, B., Aulanier, G., \& Vr{\v s}nak, B.\ 2015, \solphys, 290, 3457 
\bibitem[Schou et al.(2012)]{Schou} Schou, J., Scherrer, P.~H., Bush, R.~I., et al.\ 2012, \solphys, 275, 229

\bibitem[Schuck(2008)]{Schuck08} Schuck, P.~W.\ 2008, \apj, 683, 1134-1152


\bibitem[Shibata \& Magara(2011)]{Shibata11} Shibata, K., \& Magara, T.\ 2011, Living Reviews in Solar Physics, 8, 6 

\bibitem[Shen et al.(2012)]{Shen12} Shen, Y., Liu, Y., \& Su, J.\ 2012, \apj, 750, 12 

\bibitem[Sim{\~o}es et al.(2013)]{Simoes13} Sim{\~o}es, P.~J.~A., Fletcher, L., Hudson, H.~S., \& Russell, A.~J.~B.\ 2013, \apj, 777, 152 
\bibitem[Smyrli et al.(2010)]{Smyrli10} Smyrli, A., Zuccarello, F., Romano, P., et al.\ 2010, \aap, 521, A56 


\bibitem[Sterling \& Moore(2003)]{Sterling03} Sterling, A.~C., \& Moore, R.~L.\ 2003, \apj, 599, 1418 

\bibitem[Sturrock et al.(2015)]{Sturrock15} Sturrock, Z., Hood, A.~W., Archontis, V., \& McNeill, C.~M.\ 2015, \aap, 582, A76 

\bibitem[Song \& Zhang(2016)]{Song16} Song, Y.~L., \& Zhang, M.\ 2016, \apj, 826, 173 




\bibitem[Su et al.(2011)]{Su11} Su, J.~T., Jing, J., Wang, H.~M., et al.\ 2011, \apj, 733, 94 



\bibitem[Sudol \& Harvey(2005)]{Sudol05} Sudol, J.~J., \& Harvey, J.~W.\ 2005, \apj, 635, 647 


\bibitem[Sun et al.(2012)]{Sun12} Sun, X., Hoeksema, J.~T., Liu, Y., et al.\ 2012, \apj, 748, 77 
\bibitem[Sun et al.(2017)]{Sun17} Sun, X., Hoeksema, J.~T., Liu, Y., Kazachenko, M., \& Chen, R.\ 2017, \apj, 839, 67 

\bibitem[Suryanarayana(2010)]{Suryanarayana10} Suryanarayana, G.~S.\ 2010, \na, 15, 313 
\bibitem[Suryanarayana et al.(2015)]{Suryanarayana15} Suryanarayana, G.~S., Hiremath, K.~M., Bagare, S.~P., \& Hegde, M.\ 2015, \aap, 580, A25 


\bibitem[T{\"o}r{\"o}k et al.(2013)]{Torok13} T{\"o}r{\"o}k, T., Temmer, M., Valori, G., et al.\ 2013, \solphys, 286, 453 
\bibitem[T{\"o}r{\"o}k et al.(2014)]{Torok14} T{\"o}r{\"o}k, T., Leake, J.~E., Titov, V.~S., et al.\ 2014, \apjl, 782, L10 


\bibitem[Tziotziou et al.(2013)]{Tziotziou13} Tziotziou, K., Georgoulis, M.~K., \& Liu, Y.\ 2013, \apj, 772, 115 


\bibitem[Vemareddy et al.(2012)]{Vemareddy12} Vemareddy, P., Ambastha, A., Maurya, R.~A., \& Chae, J.\ 2012, \apj, 761, 86 
\bibitem[Vemareddy et al.(2016)]{Vemareddy16} Vemareddy, P., Cheng, X., \& Ravindra, B.\ 2016, \apj, 829, 24 

\bibitem[Wang et al.(1994)]{Wang94} Wang, H., Ewell, M.~W., Jr., Zirin, H., \& Ai, G.\ 1994, \apj, 424, 436 

\bibitem[Wang et al.(2005)]{Wang05} Wang, H., Liu, C., Deng, Y., \& Zhang, H.\ 2005, \apj, 627, 1031 
\bibitem[Wang(2006)]{Wang06} Wang, H.\ 2006, \apj, 649, 490 
\bibitem[Wang et al.(2007)]{Wang07} Wang, H., Liu, C., Jing, J., \& Yurchyshyn, V.\ 2007, \apj, 671, 973
\bibitem[Wang \& Liu(2010)]{Wang10} Wang, H., \& Liu, C.\ 2010, \apjl, 716, L195 
\bibitem[Wang et al.(2013)]{Wang13} Wang, H., Liu, C., Wang, S., et al.\ 2013, \apjl, 774, L24 

\bibitem[Wang(1996)]{Wang96} Wang, J.\ 1996, \solphys, 163, 319 
\bibitem[Wang et al.(2017)]{Wang17} Wang, J., Yan, X., Qu, Z., Xue, Z., \& Yang, L.\ 2017, \apj, 839, 128 

\bibitem[Wang et al.(2018)]{Wang18} Wang, J., Sim{\~o}es, P.~J.~A., \& Fletcher, L.\ 2018, \apj, 859, 25 

\bibitem[Wang et al.(2011)]{Wang11} Wang, P., Ding, M.-D., Ji, H.-S., \& Wang, H.-M.\ 2011, Research in As
tronomy and Astrophysics, 11, 692 

\bibitem[Wang et al.(2014b)]{Wang14b} Wang, R., Liu, Y.~D., Yang, Z., \& Hu, H.\ 2014, \apj, 791, 84 
\bibitem[Wang et al.(2016)]{Wang16} Wang, R., Liu, Y.~D., Wiegelmann, T., et al.\ 2016, \solphys, 291, 1159

\bibitem[Wang et al.(2014)]{Wang14} Wang, S., Liu, C., Deng, N., \& Wang, H.\ 2014, \apjl, 782, L31

\bibitem[Wheatland et al.(2000)]{Wheatland00} Wheatland, M.~S., Sturrock, P.~A., \& Roumeliotis, G.\ 2000, \apj, 540, 1150 


\bibitem[Wiegelmann(2004)]{Wiegelmann04} Wiegelmann, T.\ 2004, \solphys, 219, 87 

\bibitem[Xu et al.(2017)]{Xu17} Xu, Z., Jiang, Y., Yang, J., Hong, J., \& Li, H.\ 2017, \apjl, 840, L21 

\bibitem[Yamamoto et al.(2005)]{Yamamoto05} Yamamoto, T.~T., Kusano, K., Maeshiro, T., Yokoyama, T., \& Sakurai, T.\ 2005, \apj, 624, 1072 
\bibitem[Yan \& Qu(2007)]{Yan07} Yan, X.~L., \& Qu, Z.~Q.\ 2007, \aap, 468, 1083 
\bibitem[Yan et al.(2018)]{Yan18} Yan, X.~L., Wang, J.~C., Pan, G.~M., et al.\ 2018, \apj, 856, 79 
\bibitem[Zhang et al.(1994)]{Zhang94} Zhang, H., Ai, G., Yan, X., Li, W., \& Liu, Y.\ 1994, \apj, 423, 828

\bibitem[Zhang et al.(2007)]{Zang07} Zhang, J., Li, L., \& Song, Q.\ 2007, \apjl, 662, L35 

\bibitem[Zhang et al.(2008)]{Zhang08} Zhang, Y., Tan, B., \& Yan, Y.\ 2008, \apjl, 682, L133  
\bibitem[Zhang et al.(2012)]{Zhang12} Zhang, Y., Kitai, R., \& Takizawa, K.\ 2012, \apj, 751, 85 
\bibitem[Zhu et al.(2012)]{Zhu12} Zhu, C., Alexander, D., \& Tian, L.\ 2012, \solphys, 278, 121 



 \end{thebibliography}
\end{document}